\documentclass[12pt, a4paper]{article} 

\usepackage[top=25mm, bottom=25mm, left=25mm, right=25mm]{geometry}
\usepackage{titlesec}
\titleformat{\section}
  {\normalfont\Large\bfseries}{\thesection}{1em}{}
\titleformat{\subsection}
  {\normalfont\large\bfseries}{\thesubsection}{1em}{}
\titleformat{\subsubsection}
  {\normalfont\normalsize\bfseries}{\thesubsubsection}{1em}{}

\usepackage{setspace}
\usepackage{tocloft}
\usepackage{graphicx} 
\usepackage{caption}
\usepackage{subcaption} 
\usepackage{booktabs} 
\usepackage{indentfirst} 
\usepackage{newtxtext, newtxmath} 
\usepackage{bm}
\usepackage{changepage}
\usepackage{cite}
\usepackage{multirow}

\usepackage{CJKutf8}
\usepackage{xcolor}

\renewcommand{\refname}{Bibliography and References}

\begin{document}

\pagenumbering{roman}


\setcounter{page}{1} 

\vspace*{-1cm} 

\begin{center}
    {\Large \bfseries Bridging the Topological Divide: Dual-Branch Graph Distillation with Semantic Priors for Robust Cold-Start Recommendation \par}
    
    \vspace{0.8cm} 
    
    {\large Yuanfeng QIU \par}
    \vspace{0.2cm}
    {\normalsize The Hong Kong University of Science and Technology (Guangzhou) \par}
    
\end{center}

\vspace{0.5cm} 

\begin{center}
    {\large \bfseries Abstract} 
\end{center}

\vspace{-0.3cm} 

\begin{adjustwidth}{1cm}{1cm} 
\noindent
Graph Collaborative Filtering models have achieved remarkable success in recommendation systems. However, they inherently suffer from a catastrophic representation collapse in strict cold-start scenarios due to the topological isolation of unobserved items. While integrating multi-modal semantic features offers a potential remedy, directly mapping isolated semantic attributes into a globally topology-driven collaborative space frequently precipitates representation degradation and optimization conflicts. To bridge this fundamental structural divide, we propose DualGCN, an innovative Dual-Branch Graph Knowledge Distillation Framework. Operating under a decoupled two-stage training paradigm, DualGCN first constructs a robust collaborative teacher network where item representations are anchored by a topology-smoothed semantic prior, effectively reshaping the latent manifold. Subsequently, a semantic student branch employs a residual mapping mechanism and a magnitude-agnostic cosine distillation objective to achieve high-fidelity cross-modal alignment without disrupting established collaborative signals. Furthermore, we introduce a dynamic norm reconstruction mechanism to rigorously calibrate the physical scales of novel and warm items during global retrieval. Extensive experiments across three real-world benchmark datasets demonstrate that DualGCN significantly outperforms state-of-the-art baselines in strict cold-start scenarios while comprehensively preserving recommendation accuracy for highly active items.
\end{adjustwidth}

\vspace{1cm} 

\clearpage
\pagenumbering{arabic}

\section{Introduction}
Recommender systems are core components of modern online platforms, including e-commerce and social media \cite{cfapp1,cfapp2,cfapp3,cfapp4}. Among existing paradigms, Collaborative Filtering (CF) remains fundamental due to its domain-agnostic nature and proficiency in capturing user interests from historical interactions \cite{cfsurvey1, cfsurvey2, cfsurvey3, cfsurvey4}. Because these interactions naturally form bipartite graphs, Graph Collaborative Filtering (GCF) has emerged as a dominant approach \cite{gcfsurvey1, gcfsurvey2, gcfsurvey3}. Specifically, ID-based GCF models explicitly capture high-order collaborative signals via multi-hop message passing \cite{ngcf}. Architectures such as LightGCN \cite{lightgcn} and LR-GCCF \cite{LR-GCCF} further streamline this paradigm by omitting non-linear transformations, achieving high training and inference efficiency through a minimalist representation space.

However, ID-only models inherently struggle with extreme data sparsity and the cold-start problem. Without sufficient interaction history, cold-start items cannot be effectively mapped into the mature collaborative space. This creates a severe distribution shift between warm and cold items, significantly bottlenecking model performance \cite{Cold-Problem1,Cold-Problem2}. To mitigate this, a common practice is to incorporate side semantic features (e.g., text or multimodal metadata) to enrich the representation space \cite{deepmusic,dropoutnet,metaembedding,Multi-meta,CLCRec,Firzen,Unified}.

While providing valuable contextual priors to bridge sparse interaction chains\cite{CLCRec,datagood1,datagood2,datagood3}, these hybrid models typically rely on complex feature extractors and multi-layer fusion modules. This complexity not only escalates computational overhead but often compromises the purity of collaborative signals, thereby degrading general recommendation performance \cite{Gheavy1,Gheavy2}. This dilemma raises a critical question: \textit{Can we leverage the rich priors of semantic features for cold-start items while preserving the lightweight architecture and collaborative efficiency of ID-based models?
}


To answer this, we propose DualGCN, a novel dual-path graph knowledge distillation framework for cold-start recommendation. DualGCN enables approximate topological-to-semantic knowledge transfer without relying on heavy fusion modules. Our main contributions are summarized as follows:


\begin{itemize}
\item We introduce DualGCN, a lightweight distillation framework featuring an asymmetric dual-path architecture. By replacing traditional heavy fusion modules with this design, it achieves approximate  semanticknowledge transfer to the topological space, drastically reducing online inference costs.
\item We design a two-stage optimization paradigm that decouples topological updates from semantic alignment, effectively resolving gradient conflicts common in multitask graph training. Coupled with a novel topology-smoothed semantic prior, the model successfully aligns cold-start nodes within the collaborative space.
\item Extensive experiments on multiple real-world datasets demonstrate that DualGCN significantly outperforms state-of-the-art baselines in graph collaborative recommendation and achieves superior performance in cold-start scenarios.
\end{itemize}
\section{Related Work}
\subsection{Graph-based Collaborative Filtering}

Conventional Matrix Factorization (MF) models \cite{mf} primarily capture first-order interactions between users and items, often overlooking the intricate topological dependencies inherent in user-item data. To address this limitation, NGCF \cite{ngcf} introduced a message-passing mechanism via Graph Neural Networks to explicitly model high-order collaborative signals. Building on this, \cite{lightgcn} proposed a streamlined architecture by removing redundant feature transformations and non-linear activations. This simplification not only reduced computational complexity but also established LightGCN as a state-of-the-art benchmark in graph-based recommendation. However, these models remain heavily reliant on the connectivity of the interaction bipartite graph. In scenarios characterized by long-tail items or extreme data sparsity, the high-order information propagation mechanism often falters \cite{Cold-Problem2}, which fundamentally constrains the model's representation capability.
\subsection{Cold-start Recommendation}


To compensate for the scarcity of interaction signals, researchers have extensively explored leveraging item metadata, such as textual descriptions, categories, or multimodal attributes, to enhance recommendation performance. Current cold-start solutions generally bifurcate into approach-driven and data-driven strategies. Approach-driven strategies, such as DropoutNet \cite{dropoutnet}, explicitly optimize cold-start performance by randomly dropping collaborative features during training. Meta-learning models, such as MetaEmbedding \cite{metaembedding}, utilize an embedding generator to produce high-quality initial ID embeddings for new items based on their side information and a limited set of interactions. Conversely, data-driven strategies focus on augmenting the interaction matrix. For instance, content-boosted models \cite{Melville2002Content-boosted} utilize content features to generate pseudo-ratings for sparse user-item pairs before applying traditional CF. DeepMusic \cite{deepmusic} pioneered the use of deep neural networks to extract textual and acoustic features from side attributes and map them into the CF latent space, demonstrating that integrating content features effectively mitigates the cold-start dilemma. Furthermore, weakly supervised methods \cite{Alleviating} leverage user content preferences and item attributes to score new items, bridging the gap left by missing interactions.

\subsection{Representation Alignment in Recommendation}

While content features provide a viable path to alleviate cold-start issues, an inherent heterogeneous gap exists between the topology-based ID space and the attribute-based semantic space. To bridge this discrepancy, frontier research often relies on complex external graph structures for cross-view alignment. For example, DCIB \cite{DCIB} employs a dual contrastive information bottleneck to facilitate topological knowledge transfer from knowledge graphs to collaborative views. Similarly, DFHE \cite{DFHE} injects structural information into the semantic view by fusing meta-graphs and meta-paths. However, these methods depend heavily on complete prior knowledge bases, such as high-quality knowledge graphs or predefined heterogeneous information networks. In many practical and stringent industrial scenarios, such external knowledge is often unavailable\cite{semunavailable1,semunavailable2,semunavailable3}, leaving the model to rely solely on highly sparse bipartite graphs and minimal item features. Consequently, achieving efficient cross-space representation alignment, while decoupling the model from external graph structures and relying only on minimalist features, remains a critical and unresolved challenge.
\section{Preliminaries}

In this chapter, we first formalize the candidate retrieval and strict cold-start problems in recommender systems, alongside a brief review of the graph convolutional paradigm exemplified by LightGCN. Subsequently, we analyze the fundamental causes of GCN failure in cold-start scenarios from a message-passing perspective. Finally, through preliminary empirical analysis, we demonstrate the efficacy of substituting random ID embeddings with pre-trained semantic features, providing both theoretical and empirical motivation for the proposed dual-path architecture. 

\subsection{Problem Definition}

In modern recommender systems, the core objective of the candidate retrieval stage is to efficiently retrieve a highly relevant subset of items from a massive corpus for a specific user. Formally, let $\mathcal{U}$ and $\mathcal{I}$ denote the user and item sets, respectively. Based on the visibility of historical interactions, the item set is strictly partitioned into warm and cold items, such that $\mathcal{I} = \mathcal{I}_{warm} \cup \mathcal{I}_{cold}$ and $\mathcal{I}_{warm} \cap \mathcal{I}_{cold} = \emptyset$. The historical interactions between users and warm items naturally form a bipartite graph $\mathcal{G} = (\mathcal{V}_{warm}, \mathcal{E})$.

In this context, Graph Collaborative Filtering (GCF) paradigms, particularly LightGCN, have achieved remarkable success. These transductive paradigms assign learnable ID embeddings $\mathbf{e}_u, \mathbf{e}_i \in \mathbb{R}^d$ to each user $u \in \mathcal{U}$ and warm item $i \in \mathcal{I}_{warm}$ participating in training. High-order topological signals are then explicitly aggregated by performing efficient message passing on the interaction graph $\mathcal{G}$. Concurrently, items in modern systems are typically associated with rich multimodal content. Encoded by pre-trained foundation models, these multimodal attributes are formalized as initial semantic vectors $\mathbf{s}_i \in \mathbb{R}^{d_s}$.





However, traditional GCF paradigms expose critical limitations when confronting the strict cold-start (or Out-Of-Vocabulary, OOV) challenge. Specifically, when a completely novel item $i_{new} \in \mathcal{I}_{cold}$ enters the system during inference, existing models suffer a dual structural failure:
First, constrained by the transductive parameterized lookup table mechanism, the unseen $i_{new}$ lacks a pre-learned ID embedding, resulting in an immediate representation absence in the initial collaborative space. 
Second, devoid of historical interactions, its degree in the interaction graph $\mathcal{G}$ is strictly zero. This structural isolation halts the neighborhood-aggregation message-passing mechanism, completely severing the new item from the established collaborative signal space.

Consequently, the core challenge of this work is formulated as an inductive cross-modal representation alignment and latent preference prediction task. The objective is to design and jointly optimize a cross-modal mapping function $f_{\theta}: \mathbb{R}^{d_s} \rightarrow \mathbb{R}^d$, enabling the independent inference of a pseudo-collaborative representation $\mathbf{\tilde{e}}_{i_{new}} = f_{\theta}(\mathbf{s}_{i_{new}})$ for any strict cold-start item $i_{new} \in \mathcal{I}_{cold}$, relying exclusively on its semantic prior $\mathbf{s}_{i_{new}}$. To ensure the retrieval validity of $\mathbf{\tilde{e}}_{i_{new}}$, the optimization of $f_{\theta}$ is governed by two key constraints:

\begin{enumerate}
    \item \textbf{Heterogeneous Spatial Alignment:} On the visible warm subgraph $\mathcal{G}$, the mapped representation distribution $\{f_{\theta}(\mathbf{s}_i) \mid i \in \mathcal{I}_{warm}\}$ must strictly approximate the high-order collaborative space aggregated via message passing. This minimizes the heterogeneous distribution shift between the semantic space $\mathbb{R}^{d_s}$ and the collaborative space $\mathbb{R}^d$.
    
    \item \textbf{Zero-Shot Preference Fidelity:} The generated cold-start vector $\mathbf{\tilde{e}}_{i_{new}}$ must seamlessly integrate into the existing GCF retrieval space. This ensures that the similarity metric can accurately estimate the true interaction probability between a user $u \in \mathcal{U}$ and the unseen item $i_{new}$, thereby achieving zero-shot candidate retrieval under the extreme condition of zero topological degree.
\end{enumerate}

\subsection{Review of Graph Neural Networks for Recommendation}


To capture high-order collaborative signals on bipartite interaction graphs, the core paradigm of GCF focuses on optimizing initial representations via iterative message passing between nodes. Following our previous definitions, let $\mathcal{G} = (\mathcal{V}_{\text{warm}}, \mathcal{E})$ be the interaction graph comprising users and warm items. We define $\mathcal{N}_u$ and $\mathcal{N}_i$ as the first-order neighbor sets for user $u \in \mathcal{U}$ and item $i \in \mathcal{I}_{\text{warm}}$, respectively. Let $\mathbf{e}_u^{(k)}, \mathbf{e}_i^{(k)} \in \mathbb{R}^d$ denote the node embeddings generated at the $k$-th graph convolutional layer, where $k=0$ corresponds to the initialized, learnable ID embeddings.

NGCF\cite{ngcf} established the foundational paradigm for topology-driven feature evolution. It explicitly models the information interaction between a target node and its neighbors by incorporating parameterized feature transformations and non-linear activations. Specifically, the aggregation operation in the $k$-th layer of NGCF is formulated as:
\begin{equation}
\mathbf{e}_u^{(k+1)} = \sigma \left( \mathbf{W}_1 \mathbf{e}_u^{(k)} + \sum_{i \in \mathcal{N}_u} \frac{1}{\sqrt{|\mathcal{N}_u||\mathcal{N}_i|}} \left( \mathbf{W}_1 \mathbf{e}_i^{(k)} + \mathbf{W}_2 (\mathbf{e}_i^{(k)} \odot \mathbf{e}_u^{(k)}) \right) \right)
\label{eq:ngcf}
\end{equation}
Here, $\mathbf{W}_1, \mathbf{W}_2 \in \mathbb{R}^{d \times d}$ are layer-specific trainable weight matrices, $\odot$ denotes the element-wise product used to capture second-order feature interactions, and $\sigma(\cdot)$ is the non-linear activation function, typically LeakyReLU. The symmetric normalization term $\frac{1}{\sqrt{|\mathcal{N}_u||\mathcal{N}_i|}}$ is introduced to mitigate scale shifts caused by degree discrepancies.




While NGCF validates the efficacy of topological signal aggregation, subsequent research by LightGCN reveals a critical inefficiency: in an ID-only embedding space devoid of explicit semantics, complex weight matrices $\mathbf{W}$ and non-linear activations $\sigma$ are not merely redundant, but significantly exacerbate training difficulty and induce overfitting. Consequently, LightGCN\cite{lightgcn} drastically streamlines the GCF paradigm by discarding all feature transformations and non-linearities, retaining only the essential symmetric normalized neighborhood smoothing:
\begin{equation}
\mathbf{e}_u^{(k+1)} = \sum_{i \in \mathcal{N}_u} \frac{1}{\sqrt{|\mathcal{N}_u||\mathcal{N}_i|}} \mathbf{e}_i^{(k)}
\label{eq:lgnu_user}
\end{equation}
\begin{equation}
\mathbf{e}_i^{(k+1)} = \sum_{u \in \mathcal{N}_i} \frac{1}{\sqrt{|\mathcal{N}_i||\mathcal{N}_u|}} \mathbf{e}_u^{(k)}
\label{eq:lgnu_item}
\end{equation}

After executing $K$ layers of message passing, to synthesize collaborative signals across different hops and alleviate the over-smoothing problem, both NGCF and LightGCN employ a layer combination mechanism. This computes the final representations, $\tilde{\mathbf{e}}_u$ and $\tilde{\mathbf{e}}_i$, for downstream prediction by aggregating the node embeddings across all layers via a weighted sum:
\begin{equation}
\tilde{\mathbf{e}}_u = \sum_{k=0}^K \alpha_k \mathbf{e}_u^{(k)}, \quad \tilde{\mathbf{e}}_i = \sum_{k=0}^K \alpha_k \mathbf{e}_i^{(k)}
\label{eq:lgnws}
\end{equation}
where $K$ is the total number of graph convolutional layers, and the scalar $\alpha_k \geq 0$ is a hyperparameter controlling the weight of the $k$-th layer features. To maintain model simplicity and avoid additional tuning overhead, it is empirically set to a uniform distribution $\alpha_k = 1/(K+1)$ in practice. Ultimately, the model computes the predicted preference score of a user for an item via the inner product $\hat{y}_{ui} = \tilde{\mathbf{e}}_u^\top \tilde{\mathbf{e}}_i$.

Although these architectures achieve remarkable performance on active items with abundant interactions, their underlying transductive learning paradigm renders them highly vulnerable. This heavy reliance on topological connectivity inevitably leads to catastrophic performance collapse in strict cold-start scenarios. In the following subsection, we will provide concrete experimental evidence through an empirical analysis to uncover the mechanisms behind this failure.

\subsection{DualGCN's Motivation}
\subsubsection{Strict Cold-Start Breakdown in GCNs}
While LightGCN has achieved significant success on active items, the efficacy of its representation learning is inherently bottlenecked by the rigid coupling of historical interaction topology and pre-trained ID embeddings. From a matrix perspective, the smoothing propagation mechanism in LightGCN can be formulated as:
\begin{equation}
\mathbf{E}^{(k+1)} = \tilde{\mathbf{A}} \mathbf{E}^{(k)}, \quad \tilde{\mathbf{A}} = \mathbf{D}^{-1/2} \mathbf{A} \mathbf{D}^{-1/2}
\label{eq:matrix_lightgcn}
\end{equation}


where $\mathbf{E}^{(k)} \in \mathbb{R}^{(N+M) \times d}$ denotes the embedding matrix at the $k$-th layer, $\mathbf{E}^{(0)}$ represents the initial ID embeddings, and $\mathbf{A}$ is the adjacency matrix of the user-item interaction graph. However, when a novel item $i_{\text{new}} \in \mathcal{I}_{\text{cold}}$ enters the system during inference, traditional GCF models encounter a dual structural failure.


First, due to the constraints of the transductive parameterized lookup table mechanism, unseen items $i_{\text{new}}$ lack pre-learned ID embeddings. This results in an immediate representation void in the initial collaborative space, causing a collapse in model performance. Second, the absence of historical interactions implies that the degree of $i_{\text{new}}$ in the interaction graph $\mathcal{G}$ is strictly zero. Consequently, the corresponding row and column vectors in the normalized adjacency matrix $\tilde{\mathbf{A}}$ are null, leading to a structural fracture in the topology-dependent neighborhood aggregation mechanism. After $K$ layers of propagation, the final output representation of such an item degenerates into an unoptimized, isolated state:

\begin{equation}
\tilde{\mathbf{e}}_{i_{\text{new}}} = \mathbf{e}_{i_{\text{new}}}^{(0)}
\label{eq:isolated_embedding}
\end{equation}
 

This isolated feature suffers from a severe distribution gap relative to the highly clustered and mature collaborative manifold. As a result, cold-start items are significantly marginalized during vector retrieval. As illustrated in Table \ref{tab:cold_start_drop}, on the Amazon-Book2023 dataset, pure ID-driven models such as NGCF and LightGCN experience a catastrophic collapse of over 97\% in Recall@20 and NDCG@20 metrics when transitioning from warm items to strict cold-start scenarios.

\begin{table}[t]
\centering
\caption{Performance comparison on Warm vs. Strict Cold-Start scenarios on Amazon-Book2023. $\Delta \downarrow$ denotes the relative performance drop when transitioning from warm to cold-start items.}
\label{tab:cold_start_drop}
\resizebox{\linewidth}{!}{
\begin{tabular}{l|ccc|ccc}
\toprule
\multirow{2}{*}{\textbf{Method}} & \multicolumn{3}{c|}{\textbf{Recall@20}} & \multicolumn{3}{c}{\textbf{NDCG@20}} \\
\cmidrule{2-7} 
& Warm & Cold & \textbf{$\Delta \downarrow$} & Warm & Cold & \textbf{$\Delta \downarrow$} \\
\midrule
NGCF & 0.0736 & 0.0021 & \textbf{-97.1\%} & 0.0463 & 0.0009 & \textbf{-98.1\%} \\
LightGCN & 0.1244 & 0.0027 & \textbf{-97.8\%} & 0.0791 & 0.0013 & \textbf{-98.4\%} \\
\midrule
NGCF-Semantic & 0.0753 & 0.0948 & \textbf{+25.9\%} & 0.0459 & 0.0515 & \textbf{+12.2\%} \\
LightGCN-Semantic & 0.1141 & 0.0515 & \textbf{-54.9\%} & 0.0717 & 0.0270 & \textbf{-62.3\%} \\
\bottomrule
\end{tabular}
}
\end{table}


Since topological isolation severs the acquisition of collaborative signals for cold items, leveraging graph-independent prior knowledge to bridge the representation gap becomes essential. To this end, we explore the potential of substituting uninformative random IDs with high-dimensional continuous semantic features $\mathbf{s}_i \in \mathbb{R}^{d_s}$ extracted from pre-trained content models for node initialization. Specifically, on the Amazon-Book2023 dataset, we utilize the pre-trained all-mpnet-base-v2 model from the Sentence Transformers framework \cite{Galli2024Performance} as the text encoder to extract item titles and categories to generate dense semantic vectors. These vectors are subsequently reduced to the model's latent dimension via Principal Component Analysis (PCA)\cite{PCA} to serve as initial embeddings:

\begin{equation}
\mathbf{e}_i^{(0)} = \mathbf{s}_i
\label{eq:semantic_init}
\end{equation}

As shown in the lower half of Table \ref{tab:cold_start_drop}, the introduction of semantic initialization significantly mitigates or even reverses the decline in cold-start performance, providing preliminary evidence of the efficacy of semantic priors.

\begin{figure}[htbp]
    \centering
    \includegraphics[width=\textwidth]{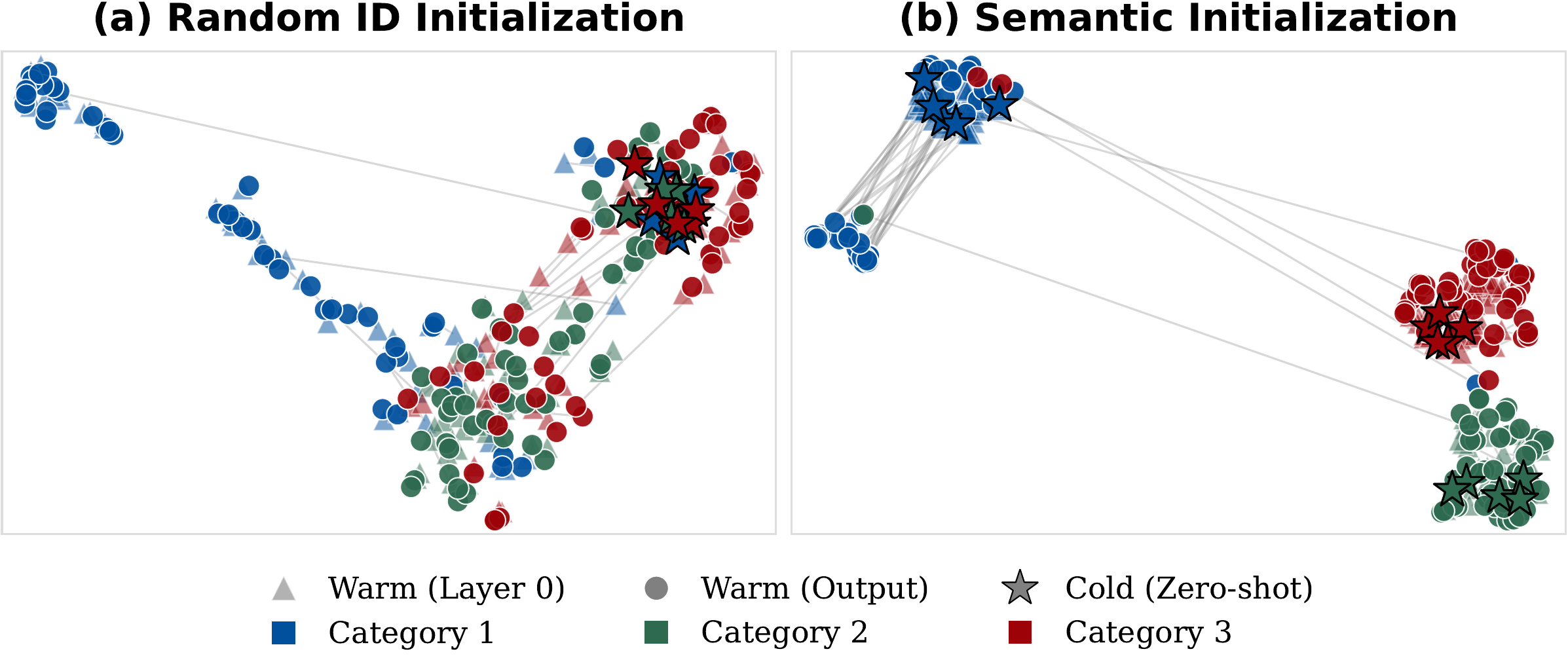}
    \caption{Uniform Manifold Approximation and Projection visualization of item representations in the latent space before and after graph message passing. With random ID initialization, the final smoothed collaborative manifold completely decouples from the initial random state, leaving degree-$0$ cold items severely isolated. In contrast, semantic initialization exerts a strong anchoring effect, preserving the topological clustering structure and inherently aligning the isolated cold items with the mature collaborative signal space.} 
    \label{fig:umap} 
\end{figure}

\subsubsection{Semantic Anchoring Effect}

To intuitively illustrate the proposed Semantic Anchoring Effect, we employ the UMAP algorithm in Figure \ref{fig:umap} to visualize the evolution of representations within the latent space. Following the message-passing mechanism of LightGCN, we extract both the initial ID embeddings and the final output representations after multi-layer aggregation and residual weighting. To construct a discriminative sample distribution, we randomly select anchor items from three distinct semantic categories and retrieve the 50 warm items with the highest semantic similarity to each anchor to form visible clusters. Furthermore, to simulate strict cold-start scenarios, we identify 5 isolated nodes with zero degree and the highest semantic proximity to the anchors for each category.


Our observations indicate that under random ID initialization, representations exhibit significant representation drift after high-order graph smoothing. Cold-start nodes, lacking topological connections, remain completely dissociated from the collaborative manifold. Conversely, the introduction of semantic vectors provides a robust anchoring effect for the nodes. For active items, the feature distribution remains tightly clustered around the initial semantic anchors even after low-pass graph filtering.

This empirical finding strongly supports the core motivation of this work: since semantic features can naturally anchor the distribution within the graph structure during the smoothing process, we can design a cross-modal mapping mechanism to align the semantic features $\mathbf{s}_{i_{\text{new}}}$ of isolated items into the mature collaborative space. This alignment allows the model to effectively overcome the cold-start dilemma caused by the failure of graph topology.

\section{Methodology}

Building on the empirical motivations and problem scope established previously, this chapter systematically delineates the proposed Dual-Branch Graph Knowledge Distillation framework (DualGCN). The framework leverages the anchoring effect of semantic features to construct asymmetric teacher and student branches, detailing their respective architectures and training strategies. Furthermore, we introduce a dynamic magnitude reconstruction mechanism and conclude with an analysis of the computational complexity of DualGCN.

\subsection{The Dual-Branch Alignment Architecture}

\subsubsection{Collaborative Teacher Branch}

The collaborative teacher branch is designed to extract the topological structures of warm-start nodes from the bipartite interaction graph $\mathcal{G} = (\mathcal{V}_{\text{warm}}, \mathcal{E})$, thereby establishing a stable and generalizable collaborative reference frame. To integrate the anchoring effect of the static semantic space into the collaborative topological space, we first define and align the physical sources of the two heterogeneous features involved. On one hand, the collaborative features $\mathbf{e}_i \in \mathbb{R}^d$ are learnable embeddings initialized via random IDs, intended to capture pure interaction preferences.


On the other hand, static semantic features $\mathbf{s}_i$ are derived from item metadata. Since cold-start items typically lack user reviews or rich interaction histories, we extract only their most fundamental attributes by concatenating the item name and category into an input description $t_i$. Specifically, we employ a pre-trained language model \texttt{all-mpnet-base-v2} \cite{Galli2024Performance}, denoted as $\Phi_{\text{enc}}(\cdot)$, to encode $t_i$ into high-dimensional dense representations. These are subsequently projected via PCA\cite{PCA} into a $d$-dimensional space aligned with the collaborative features. The specific settings for the original dimensions and $d$ are elaborated in the Experiments section. This extraction and dimensionality reduction process is formalized as:
\begin{equation}
\mathbf{s}_i = \text{PCA} \big( \Phi_{\text{enc}}(t_i) \big)
\label{eq:pca_semantic}
\end{equation}
Notably, to maintain the semantic integrity of $\mathbf{s}_i$, this semantic prior remains strictly frozen throughout the joint training process and does not participate in backpropagation updates.

However, an inherent magnitude discrepancy exists between the dense semantic vectors generated by pre-trained models and the randomly initialized collaborative vectors. To prevent this scale variance from inducing severe representation oscillation during the early stages of optimization, DualGCN first applies $L_2$ normalization to the semantic features. This strips away the absolute magnitude, extracting a pure semantic directional prior: $\bar{\mathbf{s}}_i = \frac{\mathbf{s}_i}{\|\mathbf{s}_i\|_2}$. Subsequently, we introduce a global learnable scalar $\gamma$ to adaptively regulate the injection intensity of this semantic prior. Ultimately, the 0-th layer initial embedding for a warm item $i \in \mathcal{I}_{\text{warm}}$ is formulated as a composition of the collaborative basis and the semantic anchor:
\begin{equation}
\mathbf{e}_i^{(0)} = \mathbf{e}_i + \gamma \bar{\mathbf{s}}_i
\label{eq:item_init}
\end{equation}

For a user $u \in \mathcal{U}$, the initial state retains only the ID embedding, i.e., $\mathbf{e}_u^{(0)} = \mathbf{e}_u$. During the $k$-th layer of graph convolutional propagation, nodes aggregate the $(k-1)$-th layer representations of their first-order interactive neighbors $\mathcal{N}_i$ and $\mathcal{N}_u$ via symmetric normalization:
\begin{equation}
\mathbf{e}_i^{(k)} = \sum_{u \in \mathcal{N}_i} \frac{1}{\sqrt{|\mathcal{N}_i| |\mathcal{N}_u|}} \mathbf{e}_u^{(k-1)}
\label{eq:node_aggregation}
\end{equation}

After $K$ layers of propagation, a layer combination mechanism is employed to perform mean aggregation across all layers, generating the final teacher representation $\tilde{\mathbf{e}}_i^{\mathcal{T}}$ for downstream tasks:
\begin{equation}
\tilde{\mathbf{e}}_i^{\mathcal{T}} = \frac{1}{K+1} \sum_{k=0}^K \mathbf{e}_i^{(k)}
\label{eq:teacher_rep}
\end{equation}

The final user representation $\tilde{\mathbf{e}}_u^{\mathcal{T}}$ is obtained analogously.

\textbf{Matrix Formulation Analysis:} Let $\mathbf{R} \in \mathbb{R}^{|\mathcal{U}| \times |\mathcal{I}_{\text{warm}}|}$ denote the interaction matrix between users and warm-start items. The adjacency matrix $\mathbf{A}$ of the bipartite interaction graph $\mathcal{G}$ and its symmetrically normalized form $\tilde{\mathbf{A}}$ are defined as:
\begin{equation}
\mathbf{A} = \begin{bmatrix} \mathbf{0} & \mathbf{R} \\ \mathbf{R}^T & \mathbf{0} \end{bmatrix}, \quad \tilde{\mathbf{A}} = \mathbf{D}^{-\frac{1}{2}} \mathbf{A} \mathbf{D}^{-\frac{1}{2}}
\label{eq:adj_matrix}
\end{equation}

Let $\mathbf{E}_u \in \mathbb{R}^{|\mathcal{U}| \times d}$ and $\mathbf{E}_i \in \mathbb{R}^{|\mathcal{I}_{\text{warm}}| \times d}$ define the initial ID embedding matrices for users and warm items, respectively, and $\bar{\mathbf{S}} \in \mathbb{R}^{|\mathcal{I}_{\text{warm}}| \times d}$ denote the normalized semantic matrix. The comprehensive node embedding matrix at the 0-th layer, $\mathbf{E}^{(0)}$, can be represented in block form as:
\begin{equation}
\mathbf{E}^{(0)} = \begin{bmatrix} \mathbf{E}_u \\ \mathbf{E}_i + \gamma \bar{\mathbf{S}} \end{bmatrix}
\label{eq:zero_layer_matrix}
\end{equation}

Consequently, the final global teacher feature matrix $\tilde{\mathbf{E}}^{\mathcal{T}} = \frac{1}{K+1} \sum_{k=0}^K \mathbf{E}^{(k)}$ can be linearly expanded as:
\begin{equation}
\tilde{\mathbf{E}}^{\mathcal{T}} = \underbrace{\frac{1}{K+1} \sum_{k=0}^K \tilde{\mathbf{A}}^k \begin{bmatrix} \mathbf{E}_u \\ \mathbf{E}_i \end{bmatrix}}_{\text{Graph-Propagated ID Embeddings}} + \underbrace{\gamma \frac{1}{K+1} \sum_{k=0}^K \tilde{\mathbf{A}}^k \begin{bmatrix} \mathbf{0} \\ \bar{\mathbf{S}} \end{bmatrix}}_{\text{Topology-Smoothed Semantic Prior}}
\label{eq:matrix_expansion}
\end{equation}

The second term in Eq. \ref{eq:matrix_expansion} introduces the \textit{Topology-Smoothed Semantic Prior}, where $\bar{\mathbf{S}}$ represents the initial multimodal/content features on the item side. Notably, although the user-side semantic prior is initially empty, the multi-hop information diffusion via the graph adjacency matrix $\tilde{\mathbf{A}}^k$ enables the seamless injection of item semantic features into user representations along the interaction edges.

\subsubsection{Semantic Student Branch}

To address the structural failure caused by the zero-degree isolation of cold-start items $i_{\text{new}} \in \mathcal{I}_{\text{cold}}$, the student branch relies exclusively on the static semantic prior $\bar{\mathbf{s}}_{i_{\text{new}}}$ to construct a cross-modal alignment network. Because of the inherent distribution discrepancy between the static semantic space and the dynamic collaborative space, direct mapping is highly susceptible to representation degeneration. Therefore, this branch employs an incremental prediction mechanism, using the fundamental semantic direction as an anchor to predict the feature deviation of the cold-start target within the collaborative space.

Let $\mathbf{W}_1, \mathbf{W}_2$ denote the weight matrices, $\mathbf{b}_1, \mathbf{b}_2$ the bias terms, and $\text{LN}(\cdot)$ Layer Normalization. The raw student representation $\tilde{\mathbf{e}}_i^{\mathcal{S}}$ for item $i$ is formally defined as:
\begin{equation}
\tilde{\mathbf{e}}_i^{\mathcal{S}} = \mathbf{W}_2 \sigma \left( \text{LN}(\mathbf{W}_1 \bar{\mathbf{s}}_i + \mathbf{b}_1) \right) + \mathbf{b}_2 + \gamma \bar{\mathbf{s}}_i
\label{eq:student_rep}
\end{equation}
Through residual connections, this branch learns the semantic-to-collaborative alignment mapping. As elaborated in the subsequent dynamic representation reconstruction section, cold-start items obtain an approximate collaborative projection via this semantic mapping.

\subsection{Decoupled Two-stage Optimization}

Graph neural networks inevitably suffer from the over-smoothing effect during message passing. If topological updates and semantic alignment are jointly optimized in an end-to-end framework, the dominant gradients from the highly sparse graph reconstruction can easily overwhelm the low-frequency alignment signals, thereby disrupting the inherent representations of the static semantic space. To isolate this gradient interference and ensure that the original recommendation accuracy for warm items is not compromised by the mapping network, DualGCN adopts a decoupled two-stage training paradigm.


The primary objective of the first stage is to construct a robust collaborative reference frame. During this phase, the parameters of the semantic alignment network $\{\mathbf{W}_1, \mathbf{W}_2, \mathbf{b}_1, \mathbf{b}_2\}$ are strictly frozen. The semantic features serve merely as structural anchors and do not participate in the gradient updates of the learnable embeddings. The model activates $\mathbf{e}_u, \mathbf{e}_i$, and $\gamma$, and computes the predicted score via the inner product $\hat{y}_{ui} = (\tilde{\mathbf{e}}_u^{\mathcal{T}})^T \tilde{\mathbf{e}}_i^{\mathcal{T}}$. Optimization is driven by the Bayesian Personalized Ranking (BPR) loss:
\begin{equation}
\mathcal{L}{\text{BPR}} = -\sum_{(u,i,j)} \ln \sigma \left( \hat{y}_{ui} - \hat{y}_{uj} \right) + \lambda \Theta
\label{eq:bpr_loss}
\end{equation}
where $\Theta$ represents the $L_2$ regularization term for the learnable parameters. To capture high-quality collaborative signals as effectively as possible, we employ a training strategy similar to LightGCN in this first stage: the trainable parameters are restricted solely to the 0-th layer embeddings, which makes $\Theta = \| \mathbf{E}^{(0)} \|^{2}$. In this stage we also omit the dropout mechanisms commonly used in GCN\cite{GCN} and NGCF.


The second stage aims to achieve targeted cross-spatial transfer of modal knowledge. The parameter space $\{\mathbf{e}_u, \mathbf{e}_i, \gamma\}$ converged in the first stage is permanently frozen. The final teacher representations $\tilde{\mathbf{e}}_i^{\mathcal{T}}$ serve as static soft labels to exclusively supervise the alignment network parameters. The model employs a bilaterally normalized cosine distillation loss, $\mathcal{L}_{\text{distill}}$, which discards absolute constraints in the Euclidean space and solely aligns the correlation of warm and cold features on the unit hypersphere:
\begin{equation}
\mathcal{L}_{\text{distill}} = 1 - \frac{1}{|\mathcal{I}_{\text{warm}}|} \sum_{i \in \mathcal{I}_{\text{warm}}} \frac{(\tilde{\mathbf{e}}_i^{\mathcal{S}})^T \tilde{\mathbf{e}}_i^{\mathcal{T}}}{\|\tilde{\mathbf{e}}_i^{\mathcal{S}}\|_2 \|\tilde{\mathbf{e}}_i^{\mathcal{T}}\|_2}
\label{eq:distill_loss}
\end{equation}

During the second-stage alignment, we investigated alternative optimization criteria, including Mean Squared Error (MSE) and InfoNCE\cite{InfoNCE}. However, empirical evidence indicates that imposing absolute magnitude constraints on a mapping network devoid of interaction frequency priors, or incorporating negative samples, yields no performance dividends. Consequently, we empirically adopt the cosine distillation loss as our final objective.

\subsection{Adaptive Inference and Dynamic Norm Reconstruction}

As the distillation stage decouples feature magnitudes from their directional information, directly utilizing the student branch's unit vectors for global retrieval would precipitate performance degradation due to geometric misalignment between the representation spaces. To circumvent this, we propose a dynamic norm reconstruction mechanism. Specifically, for each inference batch, the model first computes the average $L_2$ norm of the warm-start representations from the teacher branch:
\begin{equation}
\mu_{\text{norm}} = \frac{1}{|\mathcal{I}_{\text{warm}}|} \sum_{i \in \mathcal{I}_{\text{warm}}} \left\| \tilde{\mathbf{e}}_i^{\mathcal{T}} \right\|_2
\label{eq:mu_norm}
\end{equation}

To unify the downstream retrieval space, a routing mechanism is introduced to dynamically address and reconstruct the final inference representation $\mathbf{e}_i^*$. The routing logic is governed by an interaction threshold $\tau$ relative to the item's real-time degree $d_i$. For items in a cold-start or extremely sparse state ($d_i < \tau$), we extract the unit directional vector from the student network and scale it by the dynamic mean $\mu_{\text{norm}}$ to achieve scale calibration. This inference process is formally defined as:
\begin{equation}
\mathbf{e}_i^* = \begin{cases} \tilde{\mathbf{e}}_i^{\mathcal{T}}, & \text{if } d_i \ge \tau, \\ \frac{\tilde{\mathbf{e}}_i^{\mathcal{S}}}{\left\| \tilde{\mathbf{e}}_i^{\mathcal{S}} \right\|_2} \mu_{\text{norm}}, & \text{if } d_i < \tau. \end{cases}
\label{eq:reconstruction}
\end{equation}

Finally, the preference score for user $u$ toward any item $i$ is computed as the inner product $\hat{y}_{u, i} = ({\tilde{\mathbf{e}}_u^{\mathcal{T}}})^{\top} \mathbf{e}_i^*$. By aligning the physical scales of novel and established items within a global retrieval manifold, this mechanism effectively eliminates representation bias arising from uneven magnitudes without necessitating additional training overhead.

\subsection{Theoretical Validity Analysis of Cross-Space Mapping}
\begin{figure}[t]
    \centering
    \includegraphics[width=0.85\linewidth]{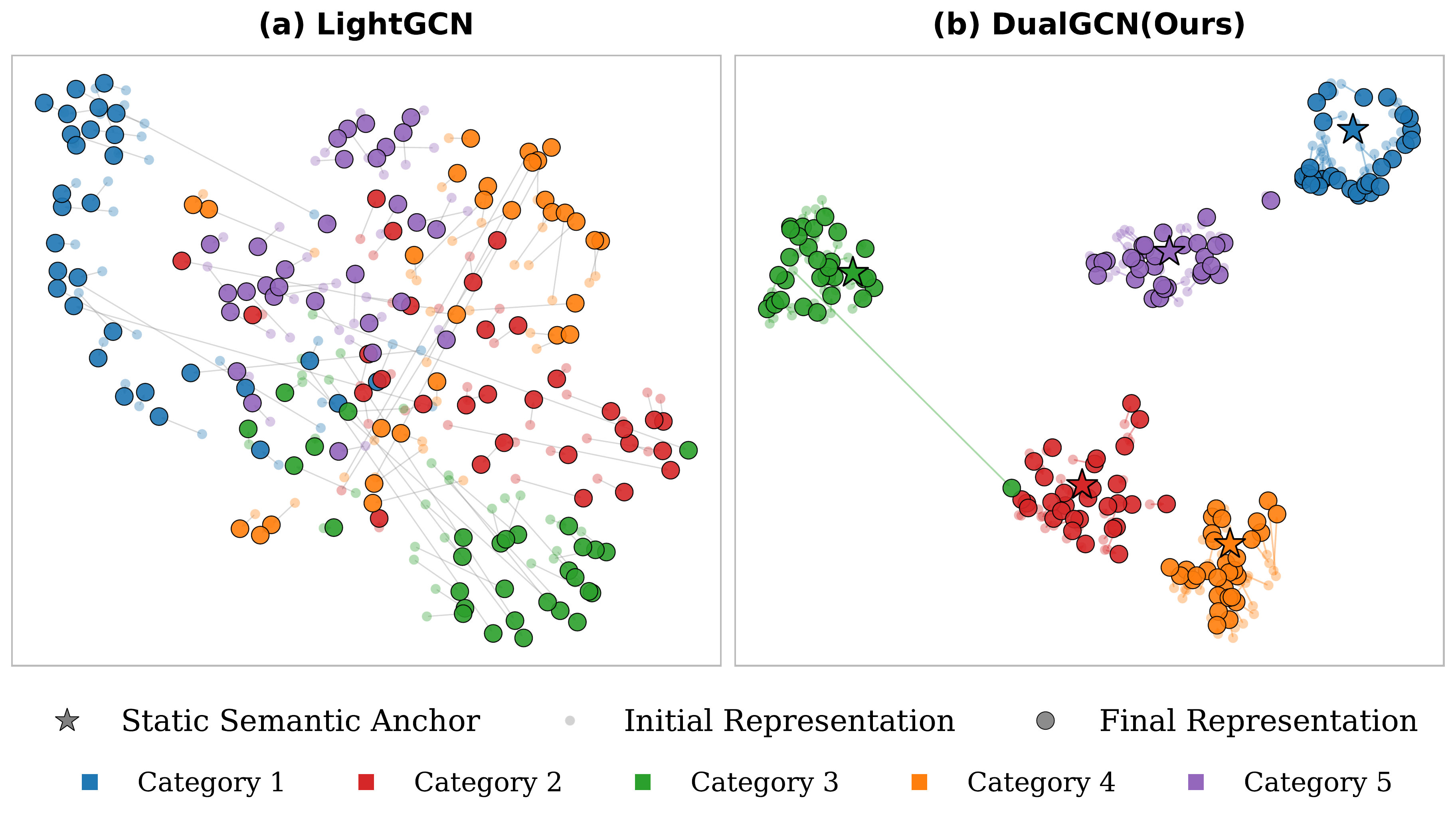}
    \caption{UMAP Visualization of feature evolution from initial to final representation. (a) LightGCN starts from chaotic random initialization. (b) DualGCN systematically expands from frozen semantic anchors $\tilde{\mathbf{E}}_{\mathrm{prior}}$ towards the global collaborative manifold.}
    \label{fig:collaborative_space}
\end{figure}



An inherent distribution gap exists between the static semantic feature space and the dynamic collaborative manifold. Consequently, compelling a mapping function to reconstruct globally topology-driven representations using solely localized and isolated semantic attributes is intrinsically unreasonable. To address this, we employ an early fusion strategy for semantic and ID representations at the input of the teacher network, introducing a Topology-Smoothed Semantic Prior:
\begin{equation}
\tilde{\mathbf{E}}_{\mathrm{prior}} = \frac{\gamma}{K+1} \sum_{k=0}^K \tilde{\mathbf{A}}^k \begin{bmatrix} \mathbf{0} \\ \bar{\mathbf{S}} \end{bmatrix}
\label{eq:semantic_prior}
\end{equation}

By freezing this prior during training, the collaborative space is reshaped into a constrained expansion manifold, anchored by semantic information and propelled by topological connectivity. To visually trace the evolution of this feature space, we sample items from five distinct categories and project their initial and final representations onto a 2D plane using UMAP. As depicted in Figure \ref{fig:collaborative_space}, the initial representations based exclusively on random IDs or local semantics exhibit a highly scattered and entangled distribution. Following the introduction of the semantic prior mechanism, the evolved representations transform from discrete or overlapping states into dense intra-class clusters. Furthermore, they maintain distinct decision boundaries across different semantic categories, effectively bridging the initial cross-modal discrepancy at a geometric level.

Empowered by this manifold alignment, the student branch is provided with robust reliability guarantees for the representation reconstruction of cold-start items. Formally, this branch employs a residual architecture, decomposing the cross-space mapping task into a superposition of a fundamental semantic anchor and a non-linear topological deviation:
\begin{equation}
\tilde{\mathbf{e}}_i^{\mathcal{S}} = \text{MLP}(\bar{\mathbf{s}}_i) + \gamma \bar{\mathbf{s}}_i
\label{eq:residual_mapping}
\end{equation}

This design fundamentally exploits the semantic anchoring property established by the teacher branch, compelling the reconstructed coordinates of cold-start items to converge within a valid joint semantic-collaborative subspace. This structural constraint ensures that even under conditions of extreme interaction sparsity, the model successfully anchors item representations within a plausible topological neighborhood, thereby preserving a robust partial order relationship for global retrieval.

\subsection{Complexity Analysis}
Despite the introduction of the dual-branch alignment architecture and the dynamic norm reconstruction mechanism, DualGCN preserves rigorous computational efficiency. The time complexity per training epoch is predominantly governed by the message-passing operations, which scale at $\mathcal{O}(K|\mathcal{E}|d)$, where $K$, $|\mathcal{E}|$, and $d$ denote the number of graph convolutional layers, the total number of interaction edges, and the embedding dimensionality, respectively. The space complexity is constrained to $\mathcal{O}(|\mathcal{V}|d + |\mathcal{E}|)$ to maintain the trainable embedding parameters and the sparse adjacency matrix. Consequently, DualGCN delivers substantial performance improvements in strict cold-start scenarios without compromising the linear scalability inherent to minimalist graph recommenders such as LightGCN.
\section{Experimental Evaluation}

In this chapter, we first detail the experimental setup. Subsequently, we conduct a comprehensive comparative evaluation against various state-of-the-art baselines under both conventional recommendation and strict cold-start scenarios. Finally, to validate the rationality and efficacy of DualGCN, we provide in-depth ablation studies and empirical analyses.

\subsection{Experimental Settings}

This section delineates the experimental setup, encompassing dataset characteristics, data preprocessing and partitioning strategies, evaluation metrics, baseline models, and specific hyperparameter configurations.

\subsubsection{Datasets}
We conduct extensive experiments on the widely recognized Amazon Reviews'23 dataset \cite{amazon}, which serves as a standard benchmark in the recommendation domain by providing both rich multimodal textual attributes and real-world interaction records. Specifically, we select three domain-specific subsets: Amazon-Books, Amazon-Electronics, and Amazon-Sports. To ensure evaluation reliability and maintain consistency with standard protocols in GCF research\cite{ngcf,lightgcn}, we apply a rigorous 10-core filtering strategy. This process eliminates users and items with fewer than 10 global interactions, thereby guaranteeing a sufficient baseline of collaborative signals within the initial feature space. Detailed statistics for each dataset are summarized in Table \ref{tab:dataset}.

\begin{table}[htbp]
  \centering
  \caption{Statistics of the datasets.}
  \label{tab:dataset}
  \begin{tabular}{lrrrr}
    \toprule
    Dataset & \#Users & \#Items & \#Interactions & Density \\
    \midrule
    Amazon-Books & 144,032 & 116,665 & 2,754,236 & 0.00016 \\
    Amazon-Electronics & 358,354 & 125,761 & 4,883,752 & 0.00011 \\
    Amazon-Sports & 25,640 & 17,543 & 313,993 & 0.00070 \\
    \bottomrule
  \end{tabular}
\end{table}

\subsubsection{Data Preprocessing and Splitting}

In the feature extraction phase, we leverage the \texttt{all-mpnet-base-v2} \cite{Galli2024Performance} encoder to map the composite textual descriptions of items, specifically their titles and categories, into 768-dimensional dense vectors. A subsequent PCA projects these high-dimensional representations into a compact 64-dimensional latent space. To establish a rigorous data partitioning scheme, we randomly isolate 10\% of the items to form the strict cold-start group. The interaction records of this subset are completely masked during the training phase to preclude any potential data leakage. The remaining 90\% of items constitute the warm-start group, and their historical interactions are randomly partitioned into training, validation, and test sets according to a 7:1:2 ratio.

\subsubsection{Evaluation Metrics}


To rigorously evaluate the ranking performance in conventional warm-start scenarios, we adopt a full-ranking strategy. For each user in the test set, the model computes preference scores and globally ranks the ground-truth interacted items against all unobserved items. Conversely, when assessing out-of-distribution cold-start performance, it is crucial to eliminate the confounding effect of popularity bias from trending items. Therefore, the candidate pool for the full ranking evaluation is strictly restricted to the pure cold-start item set.

We employ \textbf{Recall@$K$} and \textbf{NDCG@$K$} as our primary evaluation metrics. The final reported performance results are the average scores across all users in the test set. To ensure alignment with the baseline models, the truncation length is configured to $K \in \{20, 50\}$.

\subsubsection{Baselines}

To validate the superiority and efficacy of our proposed framework, we benchmark DualGCN against two representative categories of baseline methods: pure ID-based collaborative filtering models and feature-augmented cold-start models. For a fair and rigorous comparison, the semantic inputs for all feature-driven baselines are standardized to the identical 64-dimensional vectors extracted from item titles and categories.



\subsubsection*{Pure ID-based Collaborative Filtering}
\begin{itemize}
    \item \textbf{MF} \cite{mf}: As the foundational baseline for collaborative filtering, Matrix Factorization projects users and items into low-dimensional latent vectors. It relies exclusively on direct interaction histories and is optimized via the Bayesian Personalized Ranking objective.
    \item \textbf{NGCF} \cite{ngcf}: This graph-based model enriches representation learning by executing embedding propagation over the bipartite interaction graph. Such a mechanism explicitly captures the high-order connectivity inherent in the topological structure.
    \item \textbf{LightGCN} \cite{lightgcn}: By innovatively stripping away the redundant non-linear activations and feature transformations prevalent in traditional graph convolutional networks, this architecture drastically enhances both training efficiency and generalization capability. It currently stands as the state-of-the-art benchmark for pure identity-driven graph collaborative filtering.
\end{itemize}

\subsubsection*{Feature-based and Cold-start Models}
\begin{itemize}
    \item \textbf{DeepMusic} \cite{deepmusic}: This approach employs deep neural networks to extract heterogeneous modal features, such as textual and acoustic attributes, from side metadata. These extracted features are subsequently mapped into the collaborative latent space to mitigate the cold-start dilemma.
    \item \textbf{DropoutNet} \cite{dropoutnet}: Designed specifically for cold-start scenarios, this framework applies a targeted dropout mechanism to user and item identity embeddings during the training phase. This deliberate perturbation forces the model to rely on content features for preference prediction under conditions of severe interaction sparsity.
    \item \textbf{MetaEmbedding} \cite{metaembedding}: Operating as a meta-learning paradigm for cold-start recommendation, this method optimizes an embedding generator. It leverages the associated side information alongside limited interaction records of novel items to synthesize highly expressive initial identity representations.
\end{itemize}

\subsubsection{Hyperparameter Settings}

Adhering to the standard configurations established by the baseline NGCF and LightGCN models \cite{ngcf,lightgcn}, we uniformly fix the embedding dimensionality at $d = 64$ for all experiments and initialize identity embeddings via the Xavier method \cite{xavier}. The depth of the graph convolutional network is set to $K = 3$ layers. For model optimization, we employ the Adam optimizer \cite{adam} with a learning rate of $0.012$, a weight decay coefficient of $10^{-4}$, and a batch size of $262,144$. The training procedure follows a decoupled two-stage schedule. Specifically, the teacher network undergoes pre-training for $600$ epochs, followed by a knowledge distillation phase of $200$ epochs to guarantee comprehensive convergence of the model parameters at each stage.

\begin{table*}[t]
\centering
\caption{Overall Performance Comparison. The best results in each scenario are highlighted in \textbf{bold}.}
\label{tab:performance}
\resizebox{\textwidth}{!}{
\begin{tabular}{lcccccccccccc}
\toprule
\multirow{2}{*}{\textbf{Method}} & \multicolumn{4}{c}{\textbf{Books}} & \multicolumn{4}{c}{\textbf{Electronics}} & \multicolumn{4}{c}{\textbf{Sports}} \\
\cmidrule(lr){2-5} \cmidrule(lr){6-9} \cmidrule(lr){10-13}
& R@20 & R@50 & N@20 & N@50 & R@20 & R@50 & N@20 & N@50 & R@20 & R@50 & N@20 & N@50 \\
\midrule

\multicolumn{13}{c}{\textbf{Warm-Start}} \\
\midrule
MetaEmbedding & 0.0631 & 0.1028 & 0.0402 & 0.0521 & 0.0098 & 0.0180 & 0.0051 & 0.0075 & 0.0187 & 0.0380 & 0.0097 & 0.0151 \\
DropoutNet    & 0.0459 & 0.0819 & 0.0275 & 0.0382 & 0.0157 & 0.0292 & 0.0090 & 0.0128 & 0.0361 & 0.0707 & 0.0196 & 0.0291 \\
DeepMusic     & 0.0275 & 0.0484 & 0.0158 & 0.0217 & 0.0108 & 0.0190 & 0.0063 & 0.0087 & 0.0210 & 0.0409 & 0.0111 & 0.0166 \\
\midrule
MF            & 0.0063 & 0.0132 & 0.0035 & 0.0056 & 0.0103 & 0.0186 & 0.0059 & 0.0084 & 0.0058 & 0.0105 & 0.0029 & 0.0042 \\
NGCF          & 0.0736 & 0.1187 & 0.0463 & 0.0598 & 0.0173 & 0.0300 & 0.0101 & 0.0138 & 0.0371 & 0.0699 & 0.0211 & 0.0302 \\
LightGCN      & 0.1244 & 0.1986 & 0.0791 & 0.1012 & 0.0327 & 0.0533 & 0.0204 & 0.0263 & 0.0597 & 0.1024 & 0.0360 & 0.0479 \\
\midrule
DualGCN (Phase 1)& \textbf{0.1268} & \textbf{0.2032} & \textbf{0.0797} & \textbf{0.1024} & \textbf{0.0457} & \textbf{0.0746} & \textbf{0.0279} & \textbf{0.0361} & \textbf{0.0736} & \textbf{0.1267} & \textbf{0.0439} & \textbf{0.0586} \\
DualGCN (Phase 2)& 0.1262 & 0.2016 & 0.0794 & 0.1018 & 0.0456 & 0.0743 & \textbf{0.0279} & 0.0360 & 0.0717 & 0.1210 & 0.0429 & 0.0565 \\

\midrule
\multicolumn{13}{c}{\textbf{Cold-Start}} \\
\midrule
MetaEmbedding & 0.0382 & 0.0810 & 0.0193 & 0.0302 & 0.0197 & 0.0348 & 0.0106 & 0.0142 & 0.0785 & 0.1585 & 0.0379 & 0.0568 \\
DropoutNet    & 0.0780 & 0.1369 & 0.0406 & 0.0555 & 0.0225 & 0.0446 & 0.0104 & 0.0157 & 0.0926 & 0.1853 & 0.0428 & 0.0645 \\
DeepMusic     & 0.0890 & 0.1441 & 0.0489 & 0.0630 & \textbf{0.0279} & \textbf{0.0506} & \textbf{0.0138} & \textbf{0.0192} & 0.0827 & 0.1565 & 0.0393 & 0.0567 \\
\midrule
MF            & 0.0016 & 0.0041 & 0.0008 & 0.0014 & 0.0016 & 0.0039 & 0.0007 & 0.0013 & 0.0112 & 0.0297 & 0.0047 & 0.0091 \\
NGCF          & 0.0021 & 0.0052 & 0.0009 & 0.0017 & 0.0016 & 0.0040 & 0.0007 & 0.0012 & 0.0115 & 0.0279 & 0.0051 & 0.0090 \\
LightGCN      & 0.0027 & 0.0053 & 0.0013 & 0.0020 & 0.0022 & 0.0052 & 0.0010 & 0.0017 & 0.0117 & 0.0283 & 0.0049 & 0.0087 \\

\midrule
DualGCN (Phase 1)& 0.0870 & 0.1439 & 0.0478 & 0.0623 & 0.0206 & 0.0415 & 0.0096 & 0.0147 & 0.0701 & 0.1405 & 0.0320 & 0.0485 \\
DualGCN (Phase 2)& \textbf{0.1007} & \textbf{0.1666} & \textbf{0.0554} & \textbf{0.0723} & 0.0248 & 0.0492 & 0.0116 & 0.0175 & \textbf{0.1119} & \textbf{0.2180} & \textbf{0.0512} & \textbf{0.0760} \\
\bottomrule
\end{tabular}
}
\end{table*}

\subsubsection{Performance Evaluation}

This section presents a comprehensive performance comparison between the proposed DualGCN and various baseline models across both conventional and strict cold-start recommendation scenarios, followed by an in-depth analysis of its intrinsic advantages. The overall empirical results are summarized in Table \ref{tab:performance}.


\textbf{Warm-start Recommendation Performance.} We first evaluate the conventional warm-start recommendation performance on the standard test set. As demonstrated in the upper half of Table \ref{tab:performance}, the first-stage DualGCN matches or surpasses existing state-of-the-art pure ID collaborative filtering baselines, including LightGCN and NGCF, across all evaluation metrics on the three datasets. This result substantiates that integrating feature fusion and distillation mechanisms for cold-start scenarios does not compromise the model's capacity to capture and model abundant existing interaction signals. Specifically, compared to the best-performing pure ID model, LightGCN, the first-stage DualGCN achieves a remarkable improvement of 39.7\% and 23.3\% in Recall@20 on the Amazon-Electronics and Amazon-Sports datasets, respectively. Even on the Amazon-Books dataset, which features relatively denser interactions, DualGCN maintains a distinct leading advantage. This superiority is primarily attributed to the teacher network's ability to fully exploit multimodal side information, effectively bridging potential sparse connections within the pure ID interaction graph and thereby augmenting representation learning in warm-start scenarios. Concurrently, we observe that the second-stage DualGCN exhibits remarkable stability, maintaining performance highly consistent with the first stage. This confirms the efficacy of the knowledge distillation mechanism in preserving warm-start collaborative signals. Notably, while DeepMusic demonstrates competitiveness in cold-start scenarios, its performance in warm-start settings is significantly inferior to both LightGCN and DualGCN.


\textbf{Strict Cold-start Recommendation Performance. }To rigorously validate the model's true generalization capability for completely novel items, we conduct further evaluations under a strictly restricted pure cold-start scenario. As illustrated in the lower half of Table \ref{tab:performance}, given the complete absence of historical interaction records for target items, pure ID-based models reliant on graph propagation, such as LightGCN, NGCF, and MF, suffer catastrophic performance degradation with metrics plummeting to near zero. In stark contrast, DualGCN demonstrates exceptional robustness. Empirical data validates that the second-stage DualGCN effectively circumvents the cold-start failure inherent to ID embeddings, significantly outperforming traditional feature-driven solutions like DropoutNet and MetaEmbedding across multiple datasets. For instance, on the Amazon-Sports dataset, the second-stage DualGCN achieves a Recall@20 of 0.1119, yielding a 20.8\% improvement over the runner-up model, DropoutNet, while also consistently maintaining superiority across all metrics on the Amazon-Books dataset. Although DeepMusic exhibits competitive cold-start performance on Amazon-Electronics, it falters significantly in conventional warm-start evaluations when compared to LightGCN and DualGCN. This phenomenon profoundly substantiates that the decoupled two-stage training paradigm successfully distills high-order collaborative patterns inherent in complex graph structures and injects them into the side-information generator. Consequently, the model is empowered to synthesize high-quality initial collaborative representations for zero-interaction items, relying entirely on textual modality inputs.

\begin{figure}[htbp]
    \centering
    \includegraphics[width=\textwidth]{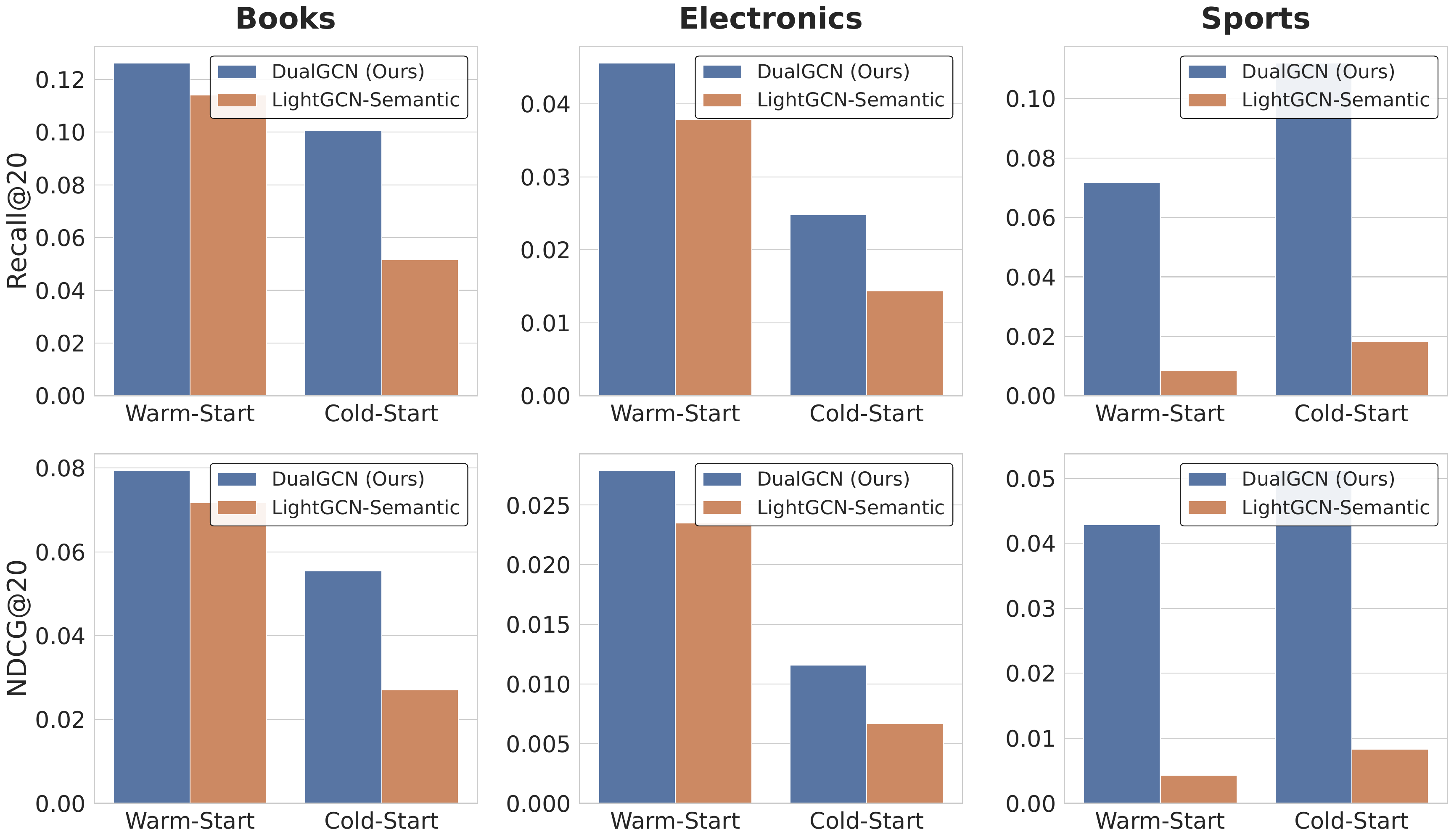}
    \caption{Ablation study on the early ID and semantic fusion strategy across different datasets and start scenarios. LightGCN-Semantic represents the variant where the randomly initialized ID embeddings in LightGCN are replaced by trainable semantic vectors.}
    \label{fig:ablation_semantic}
\end{figure}
\begin{table*}[t]
  \centering
  \caption{Ablation study of DualGCN on Warm-start and Cold-start scenarios. The best results of variants are highlighted in bold. \textbf{DualGCN} represents our default model with Cosine Loss and Semantic Residual.}
  \label{tab:ablation}
  \resizebox{\textwidth}{!}{
  \begin{tabular}{l|cccc|cccc|cccc}
    \toprule
    \multirow{2}{*}{\textbf{Architecture}} & \multicolumn{4}{c|}{\textbf{Books}} & \multicolumn{4}{c|}{\textbf{Electronics}} & \multicolumn{4}{c}{\textbf{Sports}} \\
    \cmidrule{2-13}
    & R@20 & R@50 & N@20 & N@50 & R@20 & R@50 & N@20 & N@50 & R@20 & R@50 & N@20 & N@50 \\
    \midrule
    \multicolumn{13}{c}{\textit{Part 1: Variants of Distillation Optimization}} \\
    \midrule
    \textbf{Warm-Start} & \multicolumn{12}{c}{} \\
    \quad DualGCN-MSE & \textbf{0.1264} & 0.2021 & 0.0795 & 0.1020 & \textbf{0.0456} & \textbf{0.0744} & \textbf{0.0279} & \textbf{0.0360} & 0.0716 & 0.1219 & 0.0424 & 0.0563 \\
    \quad DualGCN-InfoNCE & 0.1262 & \textbf{0.2051} & \textbf{0.0796} & \textbf{0.1022} & 0.0380 & 0.0622 & 0.0235 & 0.0304 & \textbf{0.0717} & \textbf{0.1232} & \textbf{0.0430} & \textbf{0.0571} \\
    \quad DualGCN (Default) & 0.1262 & 0.2016 & 0.0794 & 0.1018 & \textbf{0.0456} & 0.0743 & \textbf{0.0279} & \textbf{0.0360} & \textbf{0.0717} & 0.1210 & 0.0429 & 0.0565 \\
    \midrule
    \textbf{Cold-Start} & \multicolumn{12}{c}{} \\
    \quad DualGCN-MSE & 0.0951 & 0.1576 & 0.0521 & 0.0681 & 0.0236 & 0.0476 & 0.0111 & 0.0169 & \textbf{0.1127} & \textbf{0.2191} & \textbf{0.0515} & \textbf{0.0766} \\
    \quad DualGCN-InfoNCE & 0.0972 & 0.1603 & 0.0539 & 0.0701 & 0.0143 & 0.0301 & 0.0067 & 0.0105 & 0.1090 & 0.2112 & 0.0503 & 0.0744 \\
    \quad DualGCN (Default) & \textbf{0.1007} & \textbf{0.1666} & \textbf{0.0554} & \textbf{0.0723} & \textbf{0.0248} & \textbf{0.0492} & \textbf{0.0116} & \textbf{0.0175} & 0.1119 & 0.2180 & 0.0512 & 0.0760 \\
    \midrule\midrule
    \multicolumn{13}{c}{\textit{Part 2: Variants of Semantic Student Branch}} \\
    \midrule
    \textbf{Warm-Start} & \multicolumn{12}{c}{} \\
    \quad w/o Semantic Residual & 0.1262 & 0.2015 & 0.0794 & 0.1018 & 0.0456 & 0.0743 & 0.0279 & 0.0360 & 0.0717 & 0.1210 & 0.0429 & 0.0565 \\
    \quad DualGCN (Default) & 0.1262 & \textbf{0.2016} & 0.0794 & 0.1018 & 0.0456 & 0.0743 & 0.0279 & 0.0360 & 0.0717 & 0.1210 & 0.0429 & 0.0565 \\
    \midrule
    \textbf{Cold-Start} & \multicolumn{12}{c}{} \\
    \quad w/o Semantic Residual & 0.0905 & 0.1565 & 0.0447 & 0.0616 & 0.0237 & 0.0482 & 0.0105 & 0.0164 & 0.1010 & 0.2068 & 0.0408 & 0.0658 \\
    \quad DualGCN (Default) & \textbf{0.1007} & \textbf{0.1666} & \textbf{0.0554} & \textbf{0.0723} & \textbf{0.0248} & \textbf{0.0492} & \textbf{0.0116} & \textbf{0.0175} & \textbf{0.1119} & \textbf{0.2180} & \textbf{0.0512} & \textbf{0.0760} \\
    \bottomrule
  \end{tabular}
  }
\end{table*}

\subsection{Ablation Study}

In this section, we conduct comprehensive ablation studies to evaluate the individual contributions of the core components within the DualGCN framework. Specifically, we investigate the efficacy of the feature fusion mechanism, the distillation objective, and the semantic residual design by monitoring the Recall and NDCG metrics at truncation levels of 20 and 50.

\subsubsection{Effectiveness of Early ID and Semantic Fusion Strategy}

As illustrated in Figure \ref{fig:ablation_semantic}, isolating the semantic branch and subjecting it to fine-tuning precipitates a pronounced performance degradation across all datasets. This empirical observation compellingly substantiates the necessity of the anchoring effect introduced within our collaborative teacher branch. When the high-dimensional semantic features $\mathbf{s}_i$ are updated via backpropagation alongside the collaborative signals, the extreme sparsity inherent in recommendation interactions renders the model highly susceptible to overfitting. Consequently, this joint optimization severely disrupts the robust topological structure originally imbued by the large-scale pre-trained language model. Conversely, by strictly freezing $\mathbf{s}_i$, our complete DualGCN framework successfully establishes these features as a static reference frame. This mechanism effectively constrains the learning trajectory of the collaborative features $\mathbf{e}_i$, thereby yielding substantial improvements in both conventional retrieval and strict cold-start scenarios.

\subsubsection{Effect of Various Loss Functions}

To rigorously isolate the contributions of core components within DualGCN, we conduct extensive ablation studies across the Books, Electronics, and Sports datasets. As detailed in Table \ref{tab:ablation}, all model variants maintain highly consistent evaluation metrics in warm-start scenarios. For instance, the Recall@20 metric on the Books dataset uniformly stabilizes at approximately 0.126. This consistency perfectly aligns with the design rationale of our dynamic norm reconstruction mechanism, which aims to resolve the cold-start bottleneck without degrading established collaborative signals. Consequently, the subsequent analysis predominantly focuses on strict cold-start performance.

During the second-stage cross-spatial targeted transfer, we evaluate our default cosine distillation loss against alternative objectives, specifically Mean Squared Error and InfoNCE. Empirical results confirm that the default cosine formulation achieves the optimal comprehensive performance in cold-start settings. Conversely, the application of either Mean Squared Error or InfoNCE precipitates noticeable performance degradation. Specifically, substituting the objective with Mean Squared Error causes the Recall@20 on the Books and Electronics datasets to drop to 0.0951 and 0.0236, respectively, falling significantly short of the default scores of 0.1007 and 0.0248. 

The fundamental mechanism driving this phenomenon lies in the rigid constraint imposed by Mean Squared Error on absolute Euclidean distances. In graph collaborative filtering, the absolute magnitude of a node representation implicitly encodes the interaction frequency prior of the corresponding item—a prior inherently absent for cold-start items. Forcing the mapping network to approximate the absolute magnitudes of the teacher model induces magnitude distortion, thereby misleading the model into overfitting spurious popularity signals. In stark contrast, the cosine distillation adopted by DualGCN strips away these absolute constraints, executing correlation alignment exclusively on the unit hypersphere, which fundamentally enhances the model's generalization capabilities.

Furthermore, incorporating the InfoNCE objective similarly fails to yield performance dividends, even precipitating a severe decline in Recall@20 to 0.0143 on the Electronics dataset. From a mechanistic perspective, the primary objective of targeted distillation in this asymmetric framework is to achieve precise coordinate projection. Indiscriminately introducing negative samples generates extraneous repulsive forces. These forces actively disrupt the delicate geometric alignment of local features, ultimately culminating in the observed performance deterioration.

\subsubsection{Effectiveness of Semantic Residual Design}
To further validate the architectural soundness of the student branch, we evaluate a model variant devoid of the semantic residual connection, denoted as \textit{w/o Semantic Residual}. Empirical results indicate that eliminating this component precipitates a consistent and stable decline in cold-start performance across all three datasets. For instance, the Recall@20 metric drops from 0.1007 to 0.0905 on the Books dataset and from 0.1119 to 0.1010 on the Sports dataset. 

Given the inherent distribution gap between the static semantic space and the dynamic collaborative manifold, compelling the network to execute a direct, deep non-linear mapping renders it highly susceptible to representation degradation and the subsequent loss of initial semantic priors. The residual mechanism effectively circumvents this vulnerability. By establishing explicit semantic features as foundational structural anchors, this design empirically alleviates the optimization complexity inherent in cross-modal alignment. Consequently, it substantially enhances the fidelity of the approximate collaborative projections generated for cold-start items.

\section{Conclusion}

In this work, we systematically investigate the fundamental structural vulnerability of traditional graph collaborative filtering models when confronted with strict cold-start scenarios. Specifically, we identify that the topological isolation of unobserved items severs the message-passing mechanism, precipitating a catastrophic collapse in representation learning. To overcome this inherent limitation, we propose DualGCN, an innovative dual-branch graph knowledge distillation framework. By seamlessly integrating high-dimensional semantic priors extracted from pre-trained language models, we fundamentally reshape the collaborative latent space. 

The proposed framework operates through a decoupled two-stage training paradigm. Initially, the teacher branch constructs a robust, topology-smoothed collaborative reference frame anchored by semantic features. Subsequently, the student branch leverages a residual mapping network and a cosine distillation objective to achieve highly reliable cross-modal alignment. Furthermore, to unify the global retrieval space, we introduce a dynamic norm reconstruction mechanism that rigorously calibrates physical scales during inference, effectively circumventing magnitude-induced representation bias. 

Extensive empirical evaluations across three real-world benchmark datasets comprehensively validate the superiority of DualGCN. The proposed architecture not only establishes a new state-of-the-art for strict cold-start recommendation but also strictly preserves the representation capacity for conventional warm-start items without incurring substantial computational overhead. Moving forward, we intend to extend this semantic anchoring paradigm to dynamic graph environments and explore the integration of richer multi-modal signals to further enhance out-of-distribution generalization capabilities.

\clearpage

\renewcommand{\refname}{Bibliography and References}

\addcontentsline{toc}{chapter}{Bibliography and References}

\begin{singlespace}
    \bibliographystyle{ieeetr}
    
    \bibliography{references}
\end{singlespace}


\end{document}